\documentclass[aps,prb,twocolumn,groupedaddress,superscriptaddress,footinbib]{revtex4-2}

\usepackage{amsmath}
\usepackage[dvips]{color}
\usepackage[utf8]{inputenc}
\usepackage[american]{babel}
\usepackage{color}
\usepackage{graphicx}
\usepackage{epstopdf}
\usepackage{epsfig}
\usepackage{float}
\usepackage[colorlinks,linkcolor=blue,citecolor=blue,urlcolor=blue]{hyperref}
\usepackage{comment}
\usepackage{amssymb}
\usepackage{multirow}
\usepackage[table]{xcolor}
\usepackage[normalem]{ulem}

\newcommand{\angstrom}{\mbox{\normalfont\AA}}

\allowdisplaybreaks

\usepackage[normalem]{ulem}
\begin{document}

\title{Superconductivity in 2D systems enhanced by nonadiabatic phonon-production effects}

\newcommand*{\TUW}[0]{{Institute of Solid State Physics, TU Wien, 1040 Vienna, Austria}}

\newcommand*{\KACIF}[0]{{Department for Research of Materials under Extreme Conditions, Institute of Physics, 10000 Zagreb, Croatia}}

\newcommand*{\CALT}[0]{{Centre for Advanced Laser Techniques, Institute of Physics, 10000 Zagreb, Croatia}}

\author{Juraj Krsnik}
\email{jkrsnik@ifs.hr}
\affiliation{\TUW}
\affiliation{\KACIF}

\author{Dino Novko}
\email{dnovko@ifs.hr}
\affiliation{\CALT}

\author{Osor S. Bari\v{s}i\'c}
\affiliation{\KACIF}

\begin{abstract}
	
{We investigate the dynamical effects of electron-phonon coupling (EPC) on the superconducting properties of two-dimensional (2D) systems, calculating the Eliashberg function in terms of dynamically renormalized phonons. By studying different approximations for the phonon self-energy, we identify the important role of charge fluctuations in shaping the superconductivity properties, not only through the renormalization of phonon frequencies and damping rates but also through structural changes in the phonon spectral function. With the dynamical effects treated consistently, we argue that a part of the phonon spectral weight necessarily shifts to low frequencies due to the coupling to the 2D gapless plasmon. Furthermore, we find that the EPC leads to excess phonon spectral weight as well - i.e., phonon production - which generally tends to enhance the transition temperature. Our calculations point out that the influence of phonon production becomes greater as the density and the effective mass of electrons increase.}
	
\end{abstract}

\maketitle 

\textit{Introduction}.  
Most of the \textit{ab initio} studies of superconductivity properties are made under the assumption of the Born-Oppenheimer approximation\,\cite{carbotte90,giustino17}. It assumes that the fast electron subsystem accommodates instantaneously (adiabatically) to slow lattice vibrations, with no full account for the dynamical mixing between lattice and electron degrees of freedom. This framework is reasonable as long as there is a well-defined gap between the characteristic energy scales of the lattice and the electron subsystem. However, this separation of scales is not preserved in two-dimensional (2D) metals\,\cite{stern67}, when gapless plasmons emerge, potentially intersecting with phonon modes\,\cite{xiaoguang85}. Conventional phonon-mediated superconductivity in such systems can therefore be strongly influenced by a dynamical interplay between electronic excitations and phonons, requiring an equal level of treatment of electron-phonon and electron-electron correlations\,\cite{gorkov16,ruhman16}.

Extensions of \textit{ab initio} approaches aimed at incorporating the effects of dynamically screened Coulomb interaction on superconductivity were usually discussed in the context of bulk systems \cite{akashi2013,akashi2014,davydov2020,akashi2022}. Quite recently, several studies discussed the impact of nonadiabatic effects on superconductivity in layered materials using real-time time-dependent-density-functional theory combined with molecular dynamics \cite{hu2022} or by using the many-body perturbation approach together with density functional perturbation theory \cite{girotto2023}. In particular, Ref.\,\cite{girotto2023} emphasized the importance of dynamical phonon dressing via electron-hole pair excitations. However, the dynamical screening of electron-phonon matrix elements has not been accounted for, leaving the influence of phonon-plasmon coupling effects unresolved. While \textit{ab initio} techniques capable of describing plasmons, phonons, and their mixing are emerging \cite{macheda2024}, fundamental insights into the phonon-plasmon physics usually come from model calculations \cite{yokota1961,varga1965,singwi1966,kim1978,bauer09,hwang2010,caruso2018,krsnik2022}. Yet, even in these cases, we are aware of only a few attempts to study the dynamical interplay of electron-electron and electron-phonon interactions in the theory of 2D superconductivity\,\cite{ruvalds87,kresin88,bill03,veld2023}.
Such investigations are crucial for comprehending a plasmons' role in phonon-mediated pairing in layered superconductors, such as doped transition metal dichalcogenides and possibly high-$T_c$ cuprates, where a great sensitivity to dynamical screening and appearance of low-lying plasmon modes are expected\,\cite{kogar17,lin22,caruso21,bozovic90,hepting23,nag22,silkin23}. This sensitivity makes the proper treatment of screening in phonon dynamics the central issue of our work.

In this Letter, we address this issue by utilizing a 2D model-based approach and treating the electron-electron and electron-phonon interactions at the same random phase approximation (RPA) level. We compare cases with a fully dynamical screening of electron-phonon matrix elements to those with a static screening and a static polarization. 
In particular, we study the impact of dynamical phonon renormalization on the superconductivity through the phonon spectral function entering the Eliashberg function \cite{allen1974}. Our main finding is related to phonon production (PP) effects \cite{krsnik2022,krsnikPhD}, i.e., the excess of phonon spectral weight that emerges when the phonon renormalization is treated adequately. It enhances the superconductivity temperature while being strongly dependent on the phonon momentum and the adiabaticity parameter.

\begin{figure*}
	\includegraphics[width=1\textwidth]{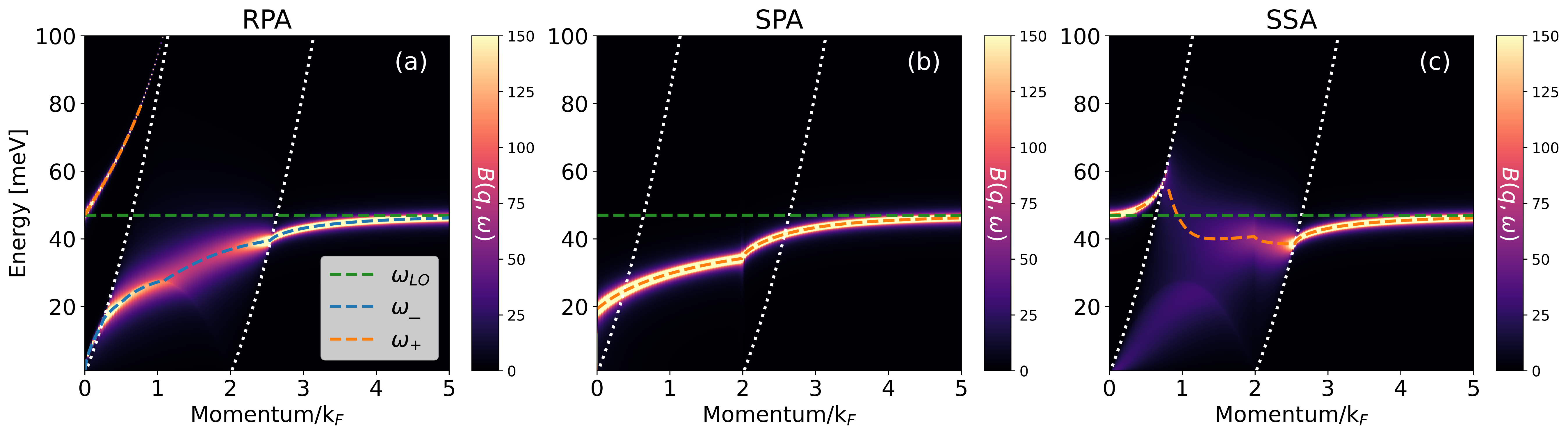}
	\renewcommand{\figurename}{Fig.}
	\caption{Phonon spectral functions for the case with $m^*/m = 0.43$ and the electron density $n=5\cdot10^{12}$ cm$^{-2}$, calculated within the (a) full dynamic RPA, (b) static polarization approximation (SPA), and (c) static screening approximation (SSA). The bare phonon energy $\hbar\omega_{LO}$ is shown with the green dashed line, while the purple and orange lines denote the renormalized phonon energies $\hbar\omega_-$ and $\hbar\omega_+$, respectively. Between the dotted white lines lies the electron-hole continuum.}
	\label{fig:phohon_spectral_functions}
\end{figure*}

\textit{Modeling}.  
We consider the standard electron-phonon model Hamiltonian given by

\begin{equation}
	\begin{split}
		\hat{H} &= \sum_{\mathbf{k}}^{}\varepsilon_{\mathbf{k}}c^\dagger_{\mathbf{k}}c_{\mathbf{k}} + \hbar\omega_{LO}\sum_{\mathbf{k}}^{}a^\dagger_{\mathbf{k}}a_{\mathbf{k}} + \frac{1}{2}\sum_{\mathbf{q}}^{}v_\mathbf{q}^{\infty}\hat{\rho}_\mathbf{q}\hat{\rho}_{-\mathbf{q}}\\
		& + \sum_{\mathbf{q}}M_\mathbf{q}\hat{\rho}_\mathbf{q}\left[  a^\dagger_{\mathbf{q}}+a_{\mathbf{-q}}\right]\;,\label{Ham}
	\end{split}	
\end{equation}
\noindent where $c^\dagger_{\mathbf{k}}$ and $a^\dagger_{\mathbf{k}}$ are the creation operators of the electron and the phonon with the wave vector $\mathbf{k}$, respectively, $\hat{\rho}_\mathbf{q} = \sum_{\mathbf{k}}c^\dagger_{\mathbf{k}+\mathbf{q}}c_{\mathbf{k}}$ is the charge density operator, and $v_\mathbf{q} =e^2/ \epsilon_0 \varepsilon_\infty q^{2}V$ the Coulomb potential screened by high-energy excitations across band gaps. We assume that the electron-phonon coupling (EPC) is dominated by the long-range polar coupling $M_\mathbf{q}$ to one longitudinal optical phonon mode with the energy $\hbar\omega_{LO} = 47$ meV. The matrix elements are given by $M_\mathbf{q} = -i\sqrt{v_\mathbf{q}^{\infty}} \sqrt{\frac{\hbar \omega_{LO}}{2}} \sqrt{1 - \frac{\varepsilon_\infty}{\varepsilon_0}}$. We fix $\varepsilon_\infty = 15.1$ and $\varepsilon_0=66.7$ \cite{laturia2018}, which corresponds to the relatively strong EPC regime with a dimensionless
EPC constant \cite{devreese2015} $\alpha\approx 0.83$. In the spirit of prototypical monolayer systems \cite{girotto2023}, we focus on a hexagonal lattice with a primitive cell $A_{pc}$ = $\frac{a^2\sqrt3}{2}$, taking $a=3.184$ \angstrom. Moreover, we assume that the relevant conduction band is weakly filled, with a quadratic electron dispersion $\varepsilon_{\mathbf{k}} = \frac{\hbar^2k^2}{2m^*}$ around the band minimum at the $\Gamma$ point, characterized with an effective mass $m^*$. This model then roughly corresponds to the case of lightly-doped transition metal dichalcogenides\,\cite{laturia2018}. Lastly, since the largest effects of the EPC within our model are present far from the Brillouin zone (BZ) boundary, we replace the momentum integrals over the full BZ with the integrals over the circle of the same area $A_{BZ}=\left( 2\pi\right)^2/ A_{pc}$.
 
We take into account the dynamical phonon renormalization through the phonon self-energy $\Pi(\mathbf{q},\omega)$. Within the RPA, for both the electron-phonon and electron-electron interactions, it reads  $\Pi(\mathbf{q},\omega) = \frac{\left| M_\mathbf{q} \right|^2}{\varepsilon(\mathbf{q},\omega)} \chi_0(\mathbf{q},\omega)$, with $\chi_0(\mathbf{q},\omega)$ and $\varepsilon (\mathbf{q},\omega)= 1-v_\mathbf{q}\chi_0(\mathbf{q},\omega)$ being the Lindhard and the dielectric function, respectively \cite{krsnik2022}. Note that in the case of 2D quadratic dispersion the Lindhard function $\chi_0(\mathbf{q},\omega)$ may be obtained analytically in a closed form \cite{mihaila2011}. In terms of $\Pi(\mathbf{q},\omega)$, the phonon spectral function is  given by
\begin{equation} \label{eq:spec_fun}
	B(\mathbf{q},\omega) = -\frac{1}{\pi}\text{Im}\left[ \frac{2\hbar\omega_{LO}}{\left( \hbar\omega\right) ^2-\left( \hbar\omega_{LO}\right) ^2-2\hbar\omega_{LO}\Pi(\mathbf{q},\omega)}\right]\;,
\end{equation}
with the corresponding renormalized phonon frequencies $\left( \hbar\omega_{\mathbf{q}}\right) ^2-\left( \hbar\omega_{LO}\right) ^2-2\hbar\omega_{LO}\Pi(\mathbf{q},\omega_{\mathbf{q}})=0$, and the phonon linewidths $\gamma_\mathbf{q} = -\text{Im}\Pi(\mathbf{q},\omega_{\mathbf{q}})$. Apart from the interaction induced $\gamma_\mathbf{q}$, we additionally assume a constant contribution $\gamma$ equal to the 1\% of the phonon energy, which may originate from an impurity, phonon-phonon or some other source of scattering.

To study the superconductivity properties, we first compute the Eliashberg function \cite{allen1974}
\begin{equation} \label{eq:eliashberg_function}
	\alpha^2F(\omega) = \frac{1}{\pi g\left( \varepsilon_{F}\right) } \sum_{\mathbf{q}}^{}\frac{\gamma_\mathbf{q} }{\hbar\omega_{\mathbf{q}}}B(\mathbf{q},\omega)\;, 
\end{equation}
where the density of states at the Fermi level for our model system reads $g\left( \varepsilon_{F}\right) = \frac{Am^*}{\hbar^2\pi}$. From $\alpha^2F(\omega)$, the cumulative EPC strength $\lambda(\omega) = 2 \int_{0}^{\omega} d\Omega \alpha^2F(\Omega)/\Omega$ is further obtained, as well as the total EPC, $\lambda= \lim\limits_{\omega\to\infty} \lambda(\omega) $. Finally, for the evaluation of $T_c$ we use the Allen-Dynes version of McMillan’s formula accounting for the effects of strong EPC~\cite{allen1975}. In all our calculations, we take for the effective Coulomb repulsion $\mu^* = 0.1$ \cite{allen1975}.

We compare three different cases discussed in the literature, each incorporating a different level of dynamics into the phonon self-energy $\Pi(\mathbf{q},\omega)$. In particular, we distinguish the fully dynamic RPA contributions, when the full frequency dependence of both $\chi_0(\mathbf{q},\omega)$ and $\varepsilon(\mathbf{q},\omega)$ is preserved; the static screening approximation (SSA), when $\varepsilon(\mathbf{q},\omega=0)$ is static, but $\chi_0(\mathbf{q},\omega)$ is still treated dynamically; and the static polarization approximation (SPA), when both $\chi_0(\mathbf{q},\omega=0)$ and $\varepsilon(\mathbf{q},\omega=0)$ are static. In this way, we can separate the influence of plasmon-phonon effects from the usually considered phonon-dressing effects by electron-hole pair excitations\,\cite{lazzeri06,caruso17,novko2020a,eiguren20,girotto2023}, since the former is present only in the fully dynamic RPA approach. By comparing in addition the SPA with the dynamical approaches, we obtain valuable insights into the overall effect of the nonadiabatic phonon renormalization on the superconductivity. 

It should be emphasized that the present considerations of the plasmon-phonon coupling effects on the superconductivity are different from the usual studies, where the impact of the plasmon and phonon modes on superconducting properties is accounted separately via the electron and phonon interaction kernels entering the Migdal-Eliashberg equations, and the dynamical screening only enters the electronic part\,\cite{ruhman16,akashi2013,akashi2014,davydov2020,kresin88,bill03}. Here, on the other hand, we inspect the often-overlooked dynamical screening of the phonon pairing mechanism, while the electronic part is accounted effectively via the static $\mu^{\ast}$.

\textit{Results}.
The effects of renormalization of the bare phonon (BP) $\hbar\omega_{LO}$ are embedded within the phonon spectral function. For $m^*/m = 0.43$ and the electron density $n=5\cdot10^{12}$ cm$^{-2}$ in the conduction band, the comparison between three different approximations is shown in Fig.~\ref{fig:phohon_spectral_functions}. Only in the fully dynamic RPA,  Fig.~\ref{fig:phohon_spectral_functions}(a), two branches $\omega_{\pm}$ appear corresponding to the plasmon-phonon coupled modes, giving two separate contributions to the Eliashberg function. Apart from the fact that the gapless branch $\omega_-$ is completely missing in both the SPA and SSA cases, the former incorrectly softens the $\omega_+$ branch in the long-wavelength limit. Note that the static screening approximation also cannot account for the linear dispersion of $\omega_+$ at small $q$ \cite{krsnikPhD}. Furthermore, the damping $\gamma_\mathbf{q}$ of the collective modes within the electron-hole continuum is absent in the purely static case. However, within the electron-hole continuum range (white dotted lines), this damping may be significant, as seen from the broadening of the peaks corresponding to $\omega_-$ in Fig.~\ref{fig:phohon_spectral_functions}(a) and $\omega_+$ in Fig.~\ref{fig:phohon_spectral_functions}(c). It is important to notice that the collective modes are more coherent in the fully dynamic RPA case in Fig.~\ref{fig:phohon_spectral_functions}(a) than in the case of the SSA in Fig.~\ref{fig:phohon_spectral_functions}(c). This emphasizes the importance of a careful treatment of dynamic effects.

	\begin{figure}
		\includegraphics[width=0.485\textwidth]{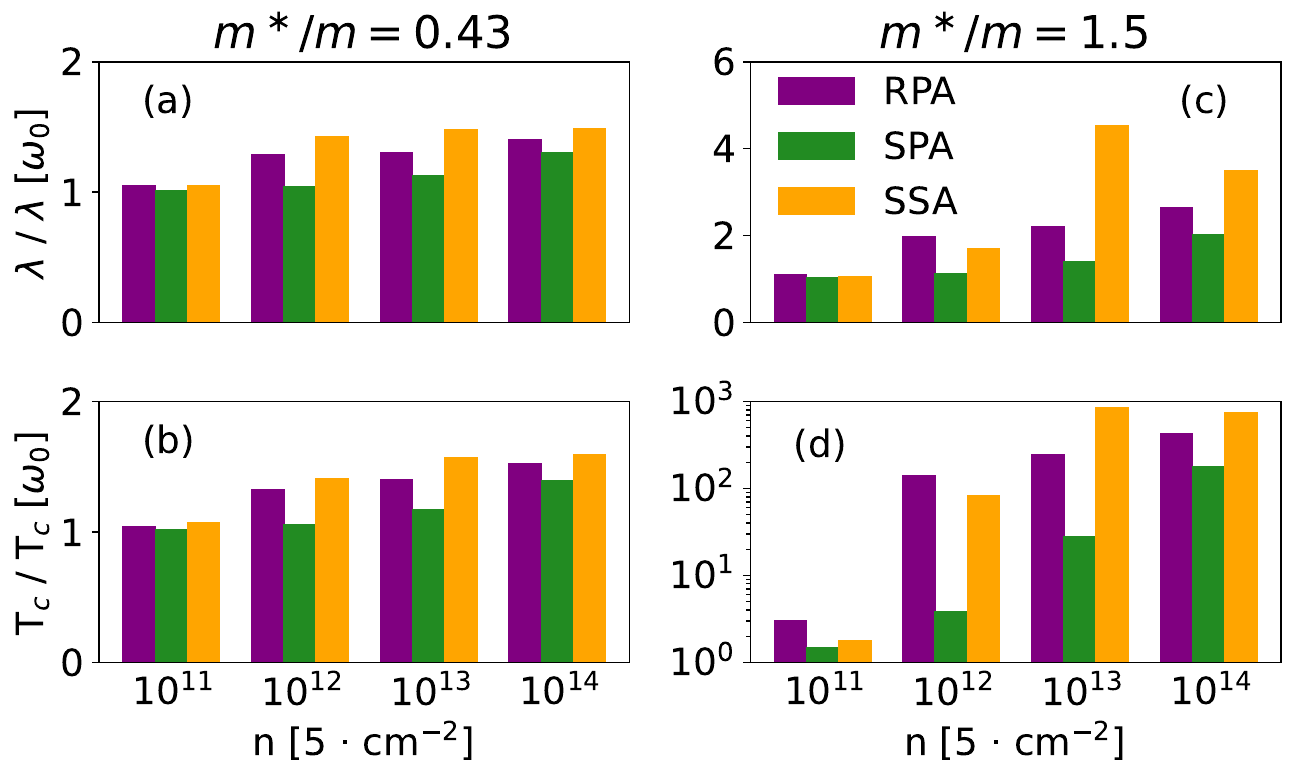}
		\renewcommand{\figurename}{Fig.}
		\caption{Electron-phonon coupling (EPC) strength (a,c) and superconducting transition temperature $T_c$ (b,d) for the cases with the effective mass $m^*/m = 0.43$ (a,b) and $m^*/m = 1.5$ (c,d) obtained within different approximations for the phonon self-energy for several electron densities. Relative values compared to the bare phonon case are shown.}
		\label{fig:lambda_tc}
	\end{figure}
 
The properties of the renormalized phonon spectral function are further inherited in the Eliashberg function $\alpha^2F(\omega)$ [see Eq.\,\eqref{eq:eliashberg_function}], the EPC strength $\lambda$, and the superconducting transition temperature $T_c$. In Figs.~\ref{fig:lambda_tc}(a,b), the latter two are shown for the $m^*/m = 0.43$ case and several different electron densities. All the results are expressed relative to the values of $\lambda\;[\omega_0]$ and $T_c\;[\omega_0]$, obtained for the BP case, i.e., when no renormalization of the phonon is taken into account, $\Pi(\mathbf{q},\omega)=0$. For the lowest density $n = 5\cdot10^{11}$ cm$^{-2}$ considered, neither adiabatic (phonon softening) nor nonadiabatic effects (mixing with the plasmon mode) significantly affect the EPC and $T_c$ compared to the BP case. This changes as the electron density increases, where we see that both the adiabatic and the nonadiabatic effects tend to increase the EPC and $T_c$ accordingly. In particular, while the SPA tends to underestimate the transition temperature when compared to the full dynamic RPA, the SSA seems to overestimate it. The maximum increase of the superconducting temperature is several tens of percent for the highest electron density $n = 5\cdot10^{14}$ cm$^{-2}$ considered.

\begin{figure}[!t]
	\includegraphics[width=.49\textwidth]{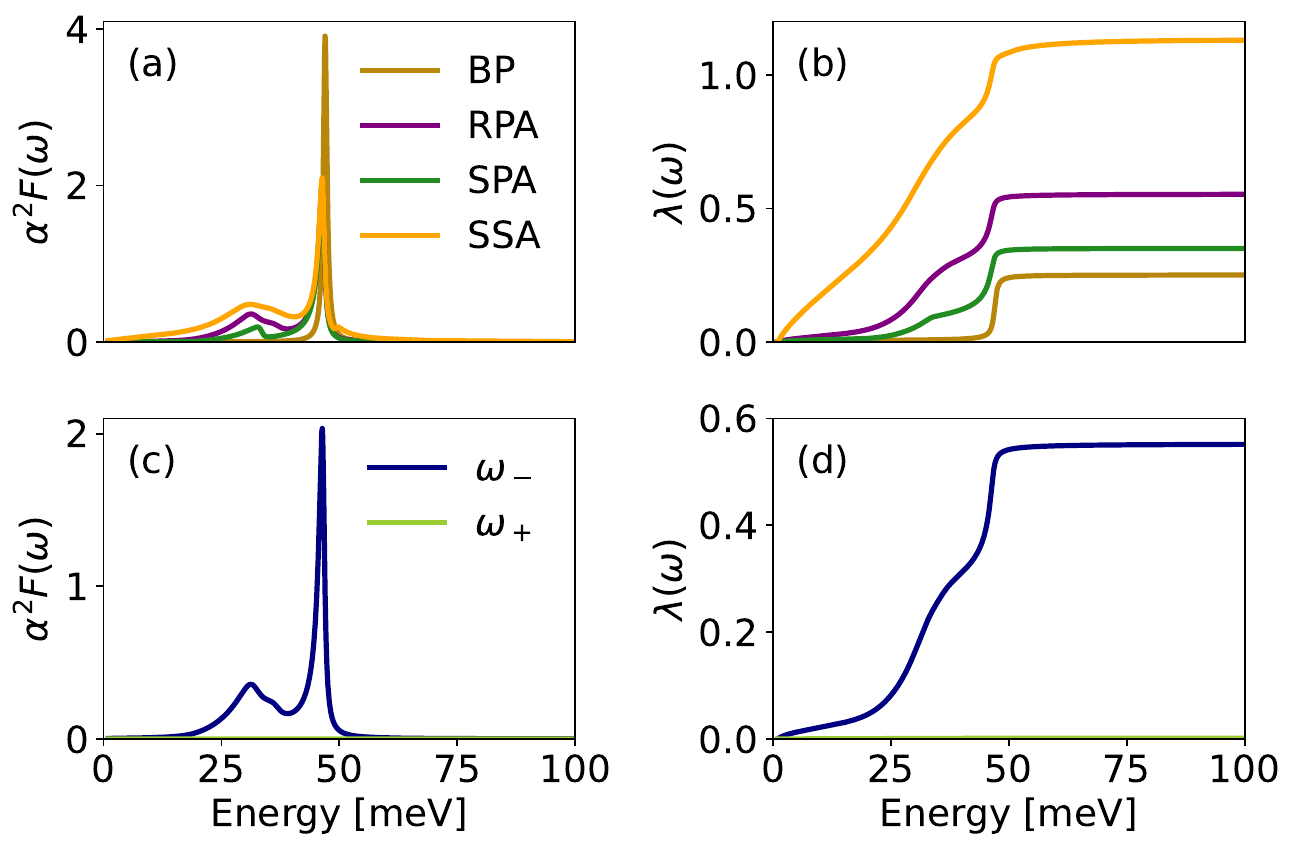}
	\renewcommand{\figurename}{Fig.}
	\caption{Eliashberg function (a,c) and cumulative electron-phonon coupling (EPC) strength (b,d) for the case with $m^*/m = 1.5$ and the electron density $n=5\cdot10^{13}$ cm$^{-2}$, calculated (a,b) within different approximations for the phonon self-energy, (c,d) for the two branches $\omega_{\pm}$ in the RPA.}
	\label{fig:eliashberg_cumulative_EPC}
\end{figure}

A further increase of the effective EPC may be achieved by reducing the steepness of the plasmon mode and the electron-hole continuum, i.e., by making the electrons heavier. For this reason, in Figs.~\ref{fig:lambda_tc}(c,d) we again consider $\lambda$ and $T_c$, but now for the case with $m^*/m = 1.5$. As expected, the phonon renormalization now has a much larger effect on both quantities, where we see that $\lambda$ might increase by several factors but $T_c$ by several orders of magnitude compared to the BP case - note logarithmic scale in Fig.~\ref{fig:lambda_tc}(d). While the increase of the electron density again results in increasing $\lambda$ and $T_c$, the rate of this increase is much larger than in the lighter electron case. Moreover, the discrepancies between different dynamic approximations become more prominent as well. Compared to the full dynamic RPA, the SSA tends to largely overestimate $T_c$, while, on the other hand, due to the nonadiabatic dynamics the SPA cannot fully account for the large increase of $T_c$. This points to the importance of the nonadiabatic effects, but also to the proper and equal level treatment of all such effects, especially as the electrons get slower.

\begin{figure}[!t]
	\includegraphics[width=.49\textwidth]{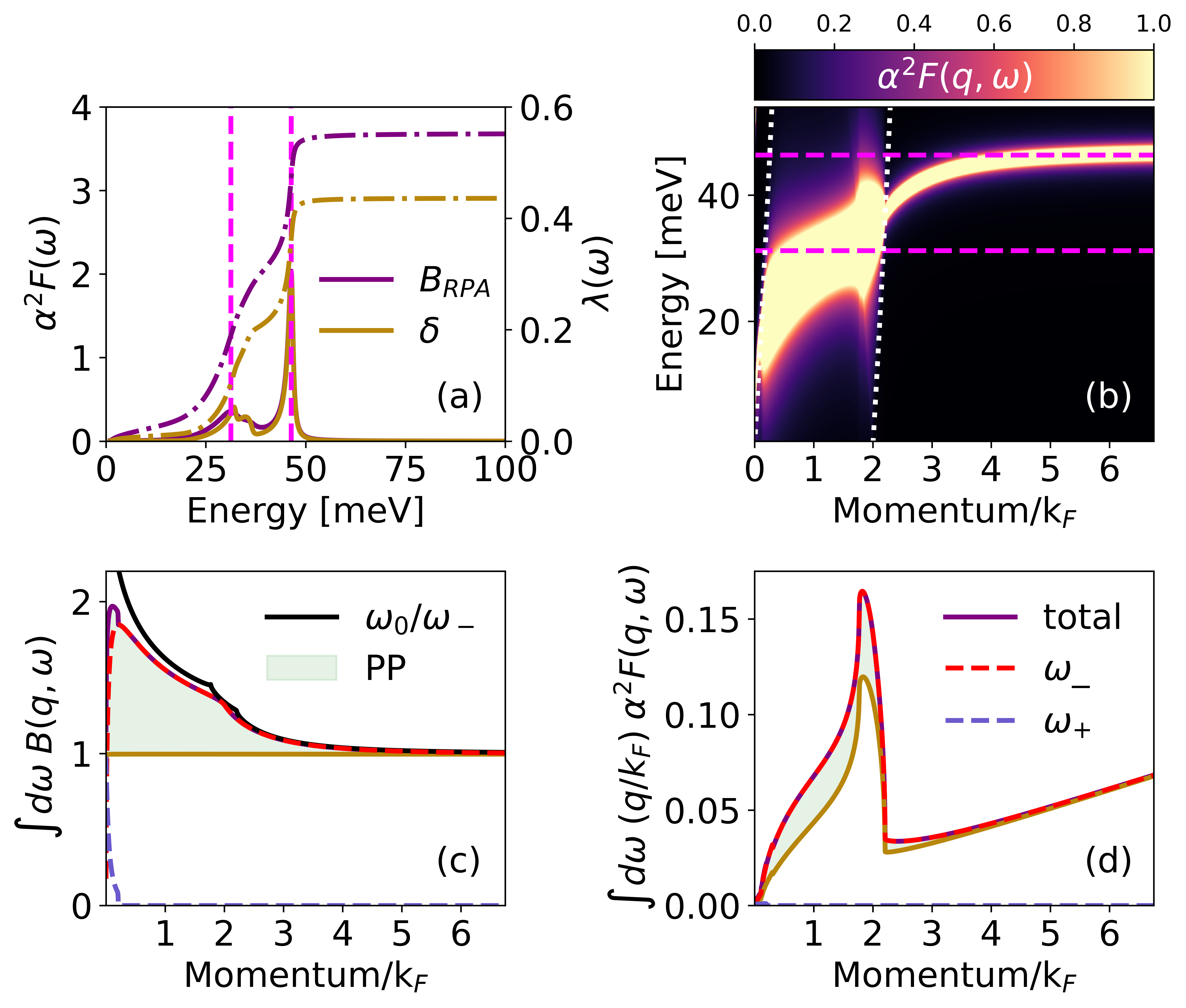}
	\renewcommand{\figurename}{Fig.}
	\caption{ (a) Eliashberg function (full lines) and cumulative EPC strength (dott-dashed lines) for the case with $m^*/m = 1.5$ and the electron density $n=5\cdot10^{13}$ cm$^{-2}$. Purple lines correspond to the case where both the renormalized frequencies, dampings, and phonon spectral function are calculated within the RPA, while orange lines to the case with the same frequencies and dampings but with the delta function at $\omega_-(\mathbf{q})$. Pink lines denote the maximums in the Eliashberg function. (b) Momentum- and frequency-resolved $\alpha^2 F(\omega)$ within the full RPA. White dotted lines denote the electron-hole continuum. (c) Momentum dependence of the phonon spectral weight. The purple solid line denotes the total, while the red and blue dashed lines the spectral weight in the $\omega_-$ and $\omega_+$ branches, respectively. The green-shaded region highlights the excess of phonon spectral weight - phonon production (PP) - in the interaction case. (d) Momentum dependence of the $\alpha^2 F(\omega)$ integrated over frequencies. The green-shaded region highlights the inherited phonon production contribution.}
	\label{fig:pp_el_qdep}
\end{figure}

To rationalize the origin of such huge enhancements of $T_c$, in Figs.~\ref{fig:eliashberg_cumulative_EPC}(a,b) we compare $\alpha^2 F(\omega)$ and the cumulative EPC strength $\lambda(\omega)$ of the bare and the renormalized phonons for $m^*/m = 1.5$ and $n=5\cdot10^{13}$ cm$^{-2}$. Since $\alpha^2 F(\omega)$ relates to the phonon density of states, for the BP case $\alpha^2 F(\omega)$ [$\lambda(\omega)$] exhibits a peak (jump) at the phonon energy $\hbar\omega_{LO}$. With the EPC included these features are still clearly present because the effect of the coupling is significant in the relatively small part of the BZ for the chosen parameter set. Nevertheless, in all the approximations considered an additional spectral weight appears below the phonon energy as well, standing behind the enhancement of the EPC and $T_c$ [note two jumps in Fig.~\ref{fig:eliashberg_cumulative_EPC}(b)]. While in the SSA case, it is clear that the additional spectral weight comes from the electron-hole continuum, for the fully dynamic RPA it is challenging to disentangle the plasmon and continuum contributions. To that end, in Figs.~\ref{fig:eliashberg_cumulative_EPC}(c,d) we additionally separate the RPA contributions coming from the two poles $\omega_\pm$.  Our results suggest that the contribution of the $\omega_+$ branch is negligible and that the $\omega_-$ branch takes all the spectral weight, where the character of the latter mode changes from plasmonic to phononic as the momentum increases (see Ref.~\cite{krsnikPhD}).

In the evaluation of $\alpha^2F(\omega)$, the phonon spectral function is throughout the literature commonly replaced by simple delta functions $\delta(\omega - \omega_\mathbf{q})$. Therefore, in Fig.~\ref{fig:pp_el_qdep}(a), we computed for comparison the Eliashberg function and the cumulative EPC, once by keeping the full dynamic RPA phonon spectral function $B_{RPA}(\mathbf{q},\omega)$, and once by replacing it with $\delta(\omega - \omega_\mathbf{q})$. In both cases, $\omega_\mathbf{q}$ and $\gamma_\mathbf{q}$ are renormalized quantities within the RPA, with the Eliashberg function exhibiting two peaks. The first is related to the phonon energy $\hbar\omega_{LO}$, while the other at lower energies piles up and exhibits a maximum within the electron-hole continuum region, see Fig.~\ref{fig:pp_el_qdep}(b). However, the frequency integration over the Eliashberg function reveals that despite the apparent similarities of $\alpha^2 F(\omega)$ in two cases, the $\lambda(\omega)$ is larger in the case with the full $B_{RPA}(\mathbf{q},\omega)$. This is due to the PP contributions \cite{krsnik2022,krsnikPhD}, which enhance the phonon spectral weight and are by construction absent in the case with the delta functions.
To that end, in Figs.~\ref{fig:pp_el_qdep}(c) and (d) the momentum-resolved phonon spectral weight and $\alpha^2 F(\omega)$ are plotted, respectively, integrated over the entire frequency range (solid purple line), from 0 to $\omega_{lim}(q)$ (dashed red line), and from $\omega_{lim}(q)$ to infinity (dashed blue line). $\omega_{lim}(q)$ is chosen such that it separates the $\omega_-$ branch from the $\omega_+$ branch \cite{omega_lim}. In particular, the phonon spectral weight integrated over frequencies reveals the presence of the PP effect that is further inherited in the Eliashberg function, with these additional spectral weights being represented by the green-shaded regions in Fig.~\ref{fig:pp_el_qdep}(c) and (d).

The momentum dependence of the PP may be used to obtain additional valuable insights into the nature of excitations\,\cite{krsnik2022,krsnikPhD}. Starting with large momenta, as the coupling to the plasmon mode becomes negligible and the phonon dispersion approaches the bare one, the PP becomes fully suppressed. As the momentum of the phonon excitation decreases, the additional phonon spectral weight becomes noticeable. For $q\gtrsim k_F$ in Fig.~\ref{fig:pp_el_qdep}(c), it should be almost completely ascribed to the adiabatic softening of the phonon modes due to the electron screening. Namely, for these momenta, the amplitude of the PP, given by the red curve in Fig.~\ref{fig:pp_el_qdep}(c), follows very closely the black curve. The latter is calculated by assuming a fully harmonic phonon mode with a softened frequency that follows the dispersion of the $\omega_-$ branch \cite{krsnik2022}. For even smaller values, i.e., $q\lesssim2k_F$, the phonon mode enters the electron-hole continuum where it becomes damped and loses its harmonic nature, exhibiting kink-like behavior. As $q$ decreases further, the amplitude of the PP starts to strongly deviate from the black curve. This is a regime when the two, basically harmonic modes strongly and nonadiabatically mix, one corresponding to the lattice vibrations and the second to the plasmon mode. This mixing results in a significant phonon spectral weight found at lower frequencies, enhancing the $\alpha^2 F(\omega)$, as seen from Fig.~\ref{fig:pp_el_qdep}(d). Note that as the electron density increases, so does the PP [e.g., see Figs. 9.9(d)-(f) in \cite{krsnikPhD}], which is in line with the increase of $\lambda$ and $T_c$ with the increase of electron density as presented in Fig.~\ref{fig:lambda_tc}. In this respect, the electron-phonon coupled systems with 2D plasmons are robust. The plasmon resonance will always drag some of the phonon spectral weight to lower frequencies.

\textit{Conclusions}.
We studied the phonon-mediated superconductivity in 2D metallic systems by treating electron-electron and electron-phonon correlations at the same dynamical RPA level. This enabled us to incorporate both the electron-hole excitations and the plasmon-phonon hybridization in the dynamical phonon renormalization. We showed that the dynamical screening of electron-phonon matrix elements is decisive in determining consistently the enhancement of the superconducting transition temperature. Distinguishing between the phonon softening, damping, and PP effects, we stressed the particular role of PP effects on the $T_c$ enhancement when these are of nonadiabatic origin due to the strong coupling with plasmon.

Our results suggest that the nonadiabatic enhancement of the superconducting transition temperature may be relevant for a variety of 2D and quasi-2D materials, hosting gapless plasmons. We predict a larger enhancement of $T_c$ for heavier electrons. Apart from 2D systems, our results could also be relevant for bulk materials that host low-energy plasmon modes, like a recently discovered interband plasmon in a boron-doped diamond \cite{bhattacharya2024}. Namely, for the boron-doped diamond the standard adiabatic theory \cite{boeri2004,giustino2007,girotto2023} predicts values of $T_c$ that are a few orders of magnitudes smaller than the experimental one \cite{ekimov2004}.

\textit{Acknowledgments}.
J.K. acknowledges support of the FWF Project No. P 36213. D.N., O.S.B., and J.K. acknowledge financial support from the Croatian Science Foundation, Grant no. UIP-2019-04-6869, IP-2022-10-9423, and IP-2022-10-3382, respectively.

\bibliography{bibliography}

\begin{thebibliography}{47}
\bibitem [{ome()}]{omega_lim}%
  \BibitemOpen
  \href@noop {} {}\bibinfo {howpublished} {To be specific, we use
  $\hbar\omega_{lim}(q) = 25$ [meV] + $250$ [meV] $q/k_F$.}\BibitemShut {Stop}%
\end{thebibliography}

\end{document}